\documentclass[preprint,aps,superscriptaddress,nofootinbib]{revtex4-1}
\usepackage{graphicx}
\usepackage[toc,page]{appendix}
\usepackage{braket}

\begin{document}
\title{Matter coupling to degenerate spacetimes in first order gravity}
 
\author{Romesh K. Kaul}
\email{kaul@imsc.res.in}
\affiliation{The Institute of Mathematical Sciences, Taramani, 
Chennai 600 113, India}

\vskip0.5cm

\begin{abstract}
We develop a systematic study of the equations of motion in the 
first order gravity with matter fields for degenerate metrics. 
Like  the Hilbert-Palatini action functional for pure gravity, 
the action  functionals  for matter fields used are first order. 
These are defined for  both invertible and non-invertible 
metrics. Description for invertible metrics is  equivalent 
to second order gravity theory with matter. For degenerate metrics
the  theory describes a different phase. The analysis for  tetrads 
with one zero eigen value in theory with scalar, Abelian vector 
gauge and fermion  matter fields is presented in detail.
 
\end{abstract}
  
\maketitle

\section{Introduction}
The usual standard description of gravity based on Einstein-Hilbert 
action functional is the second order formulation constructed 
with invertible metrics ($det~ g^{}_{\mu\nu} \ne 0$). Matter fields 
are also coupled here with action functionals which are defined for 
non-degenerate metrics. There is another theory of  gravity based 
on Hilbert-Palatini action given in terms tetrads $e^I_\mu$ and 
$SO(1,3)$ connection fields $\omega^{~IJ}_\mu$. In this description 
both the tetrads and connection fields are taken to be independent 
in the action functional. This first order formulation differs 
from the standard second order formulation in an important aspect: 
the first order theory is defined for both non-degenerate 
($det ~e^I_\mu \ne 0$) and degenerate ($det~ e^I_\mu =0$) metrics. 
For invertible tetrads, this formulation is   equivalent 
to that of the second order formulation. However, there is an additional 
phase here characterized by degenerate tetrads which has significantly 
different structure. This, therefore, provides a framework for a detail 
study of degenerate metrics. Interest in degenerate metrics has a long 
history \cite{einstein,hawking, henneaux,  tseytlin, ashtekar,  
jacobson, horowitz}. Quantum theory in first order formalism would
include contributions from configurations with both non-degenerate 
and degenerate tetrads in the functional integral. Further, 
degenerate spacetimes have also been invoked in the discussion 
of topology change \cite{horowitz, wheeler}. Topology changes  
may occur in quantum theory of spacetime. It is also possible 
that these may originate even in classical theory \cite{horowitz}.

Recently a systematic detail study of non-invertible tetrads 
configurations in first order gravity has been developed 
\cite{kaul-sandipan1, kaul-sandipan2}. For degenerate tetrads 
the theory is shown to possess solutions of vacuum equations of 
motion which generically exhibit presence of torsion. This special 
property follows even in absence of any matter fields.

The analysis in the first order gravity in \cite{kaul-sandipan1, 
kaul-sandipan2} was done without presence of any matter fields. 
To extend this to include matter fields, we need to introduce matter
field action functionals which are also first order. For a fermion 
field, the standard action functional  used is already 
first order. On the other hand,  usually used actions for other 
fields like scalar and gauge fields have second order forms and 
these are defined  only for invertible metrics. There is a straight 
forward procedure to construct first order action functionals 
from the second order actions by applying the general Ostrogradsky 
construction for lowering the number of derivatives by introducing  
additional auxiliary field variables\cite{ostro}. The first order 
matter actions so constructed then turn out to have a special 
structure which allows us to define them for both invertible and 
non-invertible tetrads. Displaying this structure explicitly, 
in the following, we shall use these action functionals for 
matter fields. For invertible metrics, these 
lead to equations of motion which are exactly equivalent to those 
obtained from the second order action functionals and hence, at
classical level, are exactly equivalent to the usual second order 
formulations. For degenerate metrics there is a different structure 
which will be studied here in detail for tetrads with one 
zero eigen value.

The article has been organized as follows. In Section II, we  
discuss coupling of a scalar matter field in first order gravity
by writing a first order action for the scalar field. The action
functional is defined for both invertible and non-invertible
tetrads. For pedagogical clarity, we explicitly demonstrate that, 
for non-degenerate tetrads, this theory is exactly equivalent to
the standard theory of scalar matter field coupled to second 
order Einstein-Hilbert gravity. 
Next we present the analysis for degenerate tetrads with one 
zero eigenvalue displaying the detail structure of the equations 
of motion. Sections III contains the analysis for first order 
gravity theory containing a vector gauge field.  First order
action functional for the vector field introduced is defined
for both non-degenerate and degenerate tetrads. For invertible 
tetrads, the theory as expected is equivalent to the second order 
gravity theory coupled to Maxwell electromagnetism. For completeness, 
we show that indeed is the case. This is followed by a detail 
analysis for degenerate tetrads. In Section  IV, we extend the 
discussion to  first order gravity with fermions presenting the 
analysis for both non-degenerate and degenerate metrics. Lastly 
Section V contains some concluding remarks.

\section{First order scalar field action}
We describe the coupling of a  scalar field in first order gravity 
through the action:
\begin{equation}\label{GS}
S~=~ S^{}_{HP\Lambda} ~+~ S^{}_{scalar}
\end{equation}  
where $S^{}_{HP\Lambda}$ is the Hilbert-Palatini action functional 
with cosmological constant ($\Lambda$) term:
\begin{equation}\label{HP}
S^{}_{HP\Lambda}~ =~ \frac{1}{8\kappa^2_{}} \int d^4x 
~\epsilon^{\mu\nu\alpha\beta}_{} ~\epsilon^{}_{IJKL}~ e^I_\mu e^J_\nu
~\left(R^{~~~KL}_{\alpha\beta}(\omega) ~-~ \frac{\Lambda}{3}~
e^K_\alpha e^L_\beta\right)
\end{equation}
and 
\begin{equation}
R^{~~~KL}_{\alpha\beta} (\omega)~=~ \partial^{}_{[\alpha} 
\omega^{~KL}_{\beta]} ~+~ \omega^{~KM}_{[\alpha} ~ 
\omega^{~~~~L}_{\beta]M}
\end{equation}
is the field strength of the $SO(1,3)$ gauge field 
$\omega^{~IJ}_{\mu}$. In this action independent fields are tetrad 
$e^I_\mu$ and connection $\omega^{~IJ}_\mu$. The  matter action 
functional  for the scalar field  coupled to the tetrad is:
\begin{equation}\label{SA}
S^{}_{scalar} ~=~ \frac{1}{6} \int d^4x~ 
\epsilon^{\mu\nu\alpha\beta}  _{} ~\epsilon^{}_{IJKL}~ 
e^J_\nu e^K_\alpha e^L_\beta \left[ 
\partial^{}_\mu \phi ~B^I_{} ~+~ \frac{1}{8}~ e^I_\mu ~\left(B^M_{} 
B^{}_M  ~-~ m^2_{} \phi^2_{} \right) \right]
\end{equation}
This action functional contains   two independent fields,  the
scalar field $\phi$ and  $B^I_{}$.

Note here the Greek indices $(\mu, ~\nu,~\alpha,~\beta)$ indicate 
the spacetime coordinates and Latin letters $(I,~J,~K,~L, ~M)$
label internal $SO(1,3)$ indices which are raised and lowered by 
the flat metric $\eta^{IJ}_{}~=~ dia ~(-1,~1,~1,~1)~ =~\eta_{IJ}^{}$.
Completely antisymmetric epsilon symbols take constant values $0$ and 
$\pm 1$ with $\epsilon^{txyz}_{}=+1$ and $\epsilon_{0123}^{}=+1$.

Like the Hilbert-Palatini action $S^{}_{HP\Lambda}$, the matter 
action functional $S^{}_{scalar}$ is first order and is defined 
for both invertible and non-invertible tetrads.  
Inverse tetrad does not appear anywhere in these 
expressions. However, as we shall see below, this matter action
functional is exactly equivalent to standard second order action 
for the scalar field $\phi$ of mass $m$ for non-degenerate tetrads.

We now obtain Euler-Lagrange equations of motion by varying the 
total action (\ref{GS}) with respect all the independent fields, 
$e^I_\mu$, $\omega^{~IJ}_\mu$, $\phi$, and $B^M_{}$. Variations with 
respect to $B_{}^M$ and $\phi$ yield respectively:
\begin{eqnarray}
&&\epsilon^{\mu\nu\alpha\beta}_{} ~\epsilon^{}_{IJKL}~ e^J_\nu
e^K_\alpha e^L_\beta \left[ \partial^{}_\mu \phi~ \delta^I_M
~+~ \frac{1}{4} ~e^I_\mu ~B^{}_M \right] ~=~0 \label{eqB}\\
&&\epsilon^{\mu\nu\alpha\beta}_{} ~\epsilon^{}_{IJKL} 
\left[ \partial^{}_\mu \left( e^J_\nu e^K_\alpha 
e^L_\beta ~ B^I_{} \right) ~+~ \frac{m^2_{}}{4} ~e^I_\mu 
e^J_\nu e^K_\alpha e^L_\beta ~\phi \right] ~=~0 \label{eqphi}
\end{eqnarray}
Next, variations of action (\ref{GS}) with respect to the connection 
field $\omega^{~IJ}_\mu$ and tetrad field $e^I_\mu$ respectively 
lead to the Euler-Lagrange equations of motion:
\begin{eqnarray}
&&\epsilon^{\mu\nu\alpha\beta}_{} ~\epsilon^{}_{IJKL}~e^K_\nu 
~D^{}_\alpha (\omega) e^L_\beta~= ~0 \label{eqomega}\\
&&\epsilon^{\mu\nu\alpha\beta}_{} ~\epsilon^{}_{IJKL}~e^J_\nu 
\left[ R^{~~~KL}_{\alpha \beta} (\omega) ~-~ \frac{2\Lambda}{3}~ 
e^K_\alpha e^L_\beta \right] ~ = ~ - 4\kappa^2_{} ~T^{~\mu}_I
\label{eqtetrad}
\end{eqnarray}
where
\begin{equation}
T^{~\mu}_I ~\equiv~ \frac{1}{2}~ \epsilon^{\mu\nu\alpha\beta}_{}
 ~\epsilon^{}_{IJKL}~ e^J_\nu e^K_\alpha \left[ \partial^{}_\beta 
 \phi~  B^L_{} ~+~ \frac{1}{6}~ e^L_\beta ~\left( B^{}_M B^M_{} 
 ~-~ m^2_{} \phi^2_{} \right) \right] \label{TS1}
\end{equation}
Here $D^{}_\mu (\omega) e^I_\nu ~\equiv~ \partial^{}_\mu e^I_\nu 
~+~ \omega^{~I}_{\mu~J} ~e^J_\nu$ is the $SO(1,3)$ gauge covariant 
derivative of the tetrad.
From Eqn.(\ref{eqtetrad}), by applying a gauge covariant derivative, 
we obtain:
\begin{equation}
 4\kappa^2_{}~D^{}_\mu(\omega) T^{~\mu}_I ~=~ - 
 ~ \epsilon^{\mu\nu\alpha\beta}_{} ~\epsilon^{}_{IJKL}~ 
 D^{}_\mu(\omega) e^J_\nu  ~R^{~~~KL}_{\alpha \beta} (\omega)
 \label{TS2}
\end{equation}
where the covariant derivative is: $D^{}_\mu(\omega)T^{~\mu}_I
~\equiv~ \partial^{}_\mu T^{~\mu}_I ~
~+~ \omega^{~~J}_{\mu I} T^{~\mu}_J$. To obtain this equation we 
have used the equation of motion (\ref{eqomega}) and also the 
Bianchi identity:
\begin{equation}
D^{}_{[\mu } (\omega) R^{~~~KL}_{\alpha \beta ]}(\omega) ~\equiv~0
\label{Bianchi}
\end{equation}

Note that, like the action functional (\ref{GS}), the Euler-Lagrange 
equations of motion (\ref{eqB}-\ref{eqtetrad}) obtained from it 
are defined for both   invertible and non-invertible tetrads.
We shall now analyze these for non-degenerate and degenerate 
tetrads separately.

\subsection{Invertible tetrads}
For the sake of completeness, for invertible tetrads, we shall 
now demonstrate that this theory is exactly same as the second 
order theory of  a scalar field coupled to gravity.

For tetrads with non-zero determinant $e ~\equiv det~e^I_\mu$,
\begin{equation} \label{e}
\epsilon^{\mu\nu\alpha\beta}_{} ~\epsilon^{}_{IJKL}~e^I_\mu 
e^J_\nu e^K_\alpha e^L_\beta ~=~ 24~e ~\ne~0,
\end{equation}
the inverse tetrad $ e^\mu_I$ is defined through relations:
\begin{eqnarray*}
e^\mu_I~e^I_\nu~=~ \delta^\mu_\nu~, ~~~~~~~~ 
 e^\mu_I~e^J_\mu~=~ \delta^J_I \nonumber
\end{eqnarray*}
For  the spacetime metric $g^{}_{\mu\nu} ~=~ e^I_\mu e^J_\nu 
~\eta_{IJ}$, the inverse is $ g^{\mu\nu}_{} ~= ~e^\mu_I e^\nu_J 
~\eta^{IJ}_{}$ and $g~\equiv ~ det ~g^{}_{\mu\nu} = -e^2_{}$.

Using Eqn.(\ref{e}) and the identity $ \epsilon^{\mu\nu\alpha\beta}_{} 
~\epsilon^{}_{IJKL}~ e^J_\nu e^K_\alpha e^L_\beta ~=~ 6e e^\mu_I$,
it is straight forward to check that the scalar action functional
(\ref{SA}) can be written as:
\begin{eqnarray}\label{SA1}
S^{}_{scalar}~=~ \int d^4x ~ e \left [ e^\mu_I~ \partial^{}_\mu 
\phi~ B^I_{} ~+~ \frac{1}{2}~ \left( B^{}_M B^M_{} ~-~m^2_{} 
~\phi^2_{} \right) \right]
\end{eqnarray}
and first order Euler-Lagrange equations of motion (\ref{eqB}) 
and (\ref{eqphi}) can be respectively recast as:
\begin{eqnarray}
&&B^{}_I ~ = ~ -~ e^\mu_I ~\partial^{}_\mu \phi \label{eqB1}\\
&&\partial^{}_\mu \left( e e^\mu_I B^I_{} \right) ~+~ e m^2_{} 
\phi~ = ~0 \label{eqphi1}
\end{eqnarray}
Eqn.(\ref{eqB1}) is a constraint reflecting the fact that $B^{}_I$ 
is not an independent field. Using this constraint in the matter
action (\ref{SA1}), we obtain
\begin{equation}
S^{}_{scalar} ~=~ -~\frac{1}{2}  \int d^4x~ e \left[ g^{\mu\nu}_{} 
\partial^{}_\mu \phi ~\partial^{}_\nu \phi ~+~ m^2_{} \phi^2_{} 
\right] \label{SA2}
\end{equation}
which is the standard second order action for a scalar field  $\phi$ 
of mass $m$ in curved spacetime. Again using the constraint 
(\ref{eqB1}) in the first order equation of motion (\ref{eqphi1}) 
leads to
\begin{equation} \label{eqphi2}
-\partial^{}_\mu \left( e g^{\mu\nu}_{} \partial^{}_\nu 
\phi \right) ~+~ e m^2_{} \phi ~=~0
\end{equation}
which is the standard second order equation of motion for  scalar 
field in curved spacetime. This equation can also be obtained 
directly by varying the second order scalar field action (\ref{SA2}) 
with respect to $\phi$.

Next, using (\ref{e}) and the identity 
$\epsilon^{\mu\nu\alpha\beta}_{} ~\epsilon^{}_{IJKL}~ e^I_\mu 
e^J_\nu  ~=~ 2e e^{[\alpha}_K e^{\beta ]}_L$, the Hilbert-Palatini
 action (\ref{HP}) can easily be seen to lead to 
\begin{equation} \label{SG1}
 S^{}_{HP \Lambda} ~=~ \frac{1}{2\kappa^2_{}}  \int d^4x ~ 
 e ~\left[~ R ~-~ 2 \Lambda ~\right]
\end{equation}
where $ R ~\equiv~ e^\mu_I e^\nu_J ~ R^{~~~IJ}_{\mu\nu} (\omega)$.

For non-degenerate tetrads, the equations of motion (\ref{eqomega}) 
is equivalent to
\begin{equation} \label{notorsion}
D^{}_{[\alpha} (\omega) ~e^I_{\beta ]} ~=~ 0
\end{equation}
This is the no-torsion condition. This equation  reflects   the fact 
that $\omega^{~IJ}_\mu$ are not independent fields and can be 
solved in terms of tetrads:
\begin{eqnarray} \label{omega(e)}
\omega^{~IJ}_\mu ~= ~\omega^{~IJ}_\mu (e) ~\equiv~ \frac{1}{2}~
\left(e^{\alpha I}_{} \partial^{}_{[\mu} e^J_{\alpha ]} 
~-~ e^{\alpha J}_{} \partial^{}_{[\mu} e^I_{\alpha ]} ~-~
e^{}_{\mu K} e^{ \alpha I}_{} e^{\beta J}_{} 
\partial^{}_{[\alpha} e^K_{\beta]} \right)  
\end{eqnarray}
Now using Eqn.(\ref{e}) and identities $ \epsilon^{\mu\nu\alpha\beta}_{} 
~\epsilon^{}_{IJKL}~ e^J_\nu ~=~ e~ e^{[\mu}_I e^\alpha_K 
e^{\beta ]}_L$ and   $ \epsilon^{\mu\nu\alpha\beta}_{} 
~\epsilon^{}_{IJKL}~ e^J_\nu  e^K_\alpha e^L_\beta ~=~ 
6e e^\mu_I $, it is straight forward to check that the last 
equation of motion (\ref{eqtetrad}) can be recast as;
\begin{equation} \label{eqtetrad1}
 R^{~\mu}_I ~-~ \frac{1}{2}~ e^\mu_I ~R ~=~ \kappa^2_{}~ 
 {\tilde T}^{~\mu}_I ~- e^\mu_I ~ \Lambda
\end{equation}
where $R^{~\mu}_I ~\equiv~ e^\alpha_I e^\beta_L e^\mu_K ~ 
R^{~~~KL}_{\alpha\beta}(\omega)$ and $R ~\equiv~ e^I_\mu ~ R^{~\mu}_I
~\equiv~ e^\alpha_I e^\beta_J ~R^{~~~IJ}_{\alpha\beta} (\omega)$ and 
$e ~{\tilde T}^{~\mu}_I ~\equiv~ T^{~\mu}_I$ with $T^{~\mu}_I$ 
of Eqn.(\ref{TS1}) given by:
\begin{eqnarray} \label{TS3}
T^{~\mu}_I ~\equiv~ e ~{\tilde T}^{~\mu}_I &=& e \left[ -~ e^\beta_I 
e^\mu_L ~ \partial^{}_\beta \phi~ B^L_{} ~+~ e^\mu_I \left( e^\beta_L
~\partial^{}_\beta \phi ~B^L ~+ ~ \frac{1}{2}~ B^{}_L B^L_{} ~-~
\frac{1}{2}~ m^2_{} \phi^2_{} \right) \right] \nonumber\\
&=& e \left[ e^\beta_I ~ \partial^{}_\beta \phi~ 
\partial^\mu_{} \phi ~- ~\frac{1}{2}~ e^\mu_I \left( \left( 
\partial \phi \right)^2_{} ~+ ~ m^2_{} \phi^2_{} \right) \right]
\end{eqnarray}
where $ \left( \partial \phi \right)^2_{} ~\equiv~ g^{\mu\nu}_{}
~\partial^{}_\mu \phi ~\partial^{}_\nu \phi$. Here we have used 
the constraint (\ref{eqB1}) in writing the last step. As expected,
varying the second order total action $S ~=~ S^{}_{HP \Lambda} 
~+~ S^{}_{scalar}$ where these two pieces of action are as in 
Eqns. (\ref{SG1}) and (\ref{SA2}) with  connection field  as in 
(\ref{omega(e)}), with respect to the tetrad $e^I_\mu$  directly
yields second order equation of motion  (\ref{eqtetrad1}).
 
Lastly, for invertible tetrads,  we notice that Eqn.(\ref{TS2}) 
leads to the conservation equation:
\begin{equation}\label{TS4}
D^{}_\mu (\omega) T^{~\mu}_I \equiv ~ D^{}_\mu(\omega)
\left (e~ {\tilde T}^{~\mu}_I \right) ~=~ 0
\end{equation}
where we have used the no-torsion condition (\ref{notorsion}).   
 
For invertible tetrads where the connection fields are given in 
terms of tetrad fields as in Eqn.(\ref{omega(e)}), the local $SO(1,3)$ 
field strength $R^{~~~IJ}_{\mu\nu} (\omega)$ is related to the 
Riemann curvature $R^{~~~~\rho}_{\mu\nu\lambda} (\Gamma)$ written 
in terms of Christoffel symbol $\Gamma^{~~\lambda}_{\mu\nu}$  as:
\begin{eqnarray*}
R^{~~~~\rho}_{\mu\nu\lambda} (\Gamma) ~=~ e^{}_{\lambda I} e^\rho_J
~R^{~~~IJ}_{\mu\nu} (\omega)
\end{eqnarray*}
With this we clearly notice that equations in (\ref{eqtetrad1}) 
are the standard second order Einstein field equations for gravity 
with scalar matter. Note that ${\tilde T}^{\mu\nu}_{} ~\equiv~
e^{\mu I}_{} {\tilde T}^{~\nu}_I$ constructed from (\ref{TS3}) 
is the standard  stress-energy tensor for the  scalar field 
matter with Eqn.(\ref{TS4}) as the conservation condition for the 
stress-energy tensor. Thus first order theory is exactly equivalent 
to the standard theory of gravity for invertible tetrads. However, 
first order theory has an additional phase containing solutions 
with degenerate tetrads.
 
\subsection{Non-invertible tetrads}
We shall now study the case where the tetrad $e^I_\mu$ has one zero
eigenvalue. We parameterize this tetrad as:
\begin{eqnarray}\label{tetradD}
e_\mu^I~=~\left(\begin{array}{ccc}
0 & 0\\
0 & e^i_a\end{array}\right) 
\end{eqnarray}
where $e^I_t ~=~e^0_a ~=~ 0$ and the $3\times 3$ block of triads 
$e^i_a ~ ~( i=1,2,3; ~a=x, y,z)$ is invertible with $det~e^i_a 
~\equiv~ {\hat e} ~\ne ~ 0$. Inverse triad will be denoted by 
${\hat e}^a_i : ~~{\hat e}^a_i e^j_a ~= \delta_i^j, ~ {
\hat e}^a_i e^i_b ~=~ \delta^a_b$. The degenerate metric is:
\begin{eqnarray*} 
g^{}_{\mu\nu} ~=~ e^I_\mu ~e^J_\nu ~\eta^{}_{IJ}
~=~\left(\begin{array}{ccc}
0 & 0\\
0 & g^{}_{ab}\end{array}\right); ~~~~ g^{}_{ab} ~=~ e^i_a e^i_b 
\end{eqnarray*}

Let us now analyze the Euler-Lagrange equations of motion 
(\ref{eqB}-\ref{eqtetrad}) for the degenerate tetrad (\ref{tetradD}). 
The matter Euler-Lagrange equations of motion (\ref{eqB}) and
(\ref{eqphi}) lead to:
\begin{equation}\label{eqBphi}
\partial^{}_t \phi ~=~ 0 ; ~~~~~~ \partial^{}_t \left( 
{\hat e} B^0_{}\right) ~=~0
\end{equation}
where we have used the identity $ 6 {\hat e} ~=~ \epsilon^{abc}_{}
~\epsilon^{}_{ijk}~ e^i_a e^j_b e^k_c$. Thus this set of 
Euler-Lagrange equations of motion make the scalar field $\phi$ 
and $({\hat e}~ B^0_{})$ time independent. Note that there are
no constraints on $\partial^{}_a \phi$ and $ B^i_{}$.

Next we shall study  Euler-Lagrange equations of motion (\ref{eqomega})
and (\ref{eqtetrad}) following closely the discussions in 
\cite{kaul-sandipan1}. For degenerate tetrad (\ref{tetradD}),
twenty four equations of motion in (\ref{eqomega}) can be broken into
four sets of $3$, $3$, $9$, and $9$ equations respectively as follows:
\begin{eqnarray}
\epsilon^{abc}_{}~ \epsilon^{}_{jkl}~e^k_a D^{}_b(\omega) e^l_c &=&~0 
\label{eqomega1}\\
\epsilon^{abc}_{}~ \epsilon^{}_{ijk}~e^k_a D^{}_b(\omega) e^0_c &=&~0 
\label{eqomega2}\\
\epsilon^{abc}_{}~ \epsilon^{}_{ijk}~e^j_b D^{}_t(\omega) e^k_c &=&~0 
\label{eqomega3}\\
\epsilon^{abc}_{}~ \epsilon^{}_{ijk}~e^j_b D^{}_t(\omega) e^0_c &=&~0
\label{eqomega4}  
\end{eqnarray}
The last equation (\ref{eqomega4}) is solved by
\begin{equation}\label{solomega1}
 \omega^{~0i}_t ~=~0
\end{equation}
Next, Eqn.(\ref{eqomega3}) implies that $D^{}_t(\omega)e^i_a~\equiv~
\partial^{}_t e^i_a~+~\omega^{~ij}_t e^j_a~=~0$, which can be 
solved for $\omega^{~ij}_t$ as:
$~\omega^{~ij}_t = {\hat e}^a_i \partial^{}_t e^j_a
= e^i_a \partial^{}_t {\hat e}^a_j = -~{\hat e}^a_j 
\partial^{}_t e^i_a = -~ e^j_a \partial^{}_t {\hat e}^a_i $.
Note that $\partial^{}_t g^{}_{ab} ~ \equiv~ D^{}_t(\omega) e^i_a~
e^i_b ~+~ e^i_a D^{}_t (\omega) e^i_b ~=0$. This implies that  
$t$ dependence of the triad fields $e^i_a$ is only a gauge 
artifact and hence can be rotated away completely by an internal 
space $O(3)$ rotation. Thus we make a gauge choice such that
\begin{equation}
\partial^{}_t e^i_a ~=~0~  ~~~{\rm and} ~~~{\rm hence} ~~~ 
\omega^{~ij}_t ~=~0 \label{solomega2}
 \end{equation}
Eqn.(\ref{eqomega2}) can be solved by:
\begin{equation}
\omega^{~0j}_a ~\equiv~ M^{~j}_a ~=~e^i_a ~ M^{ij}_{} ~~~~~ {\rm with} 
~~~~~ M^{ij}_{}~=~ M^{ji}_{} \label{solomega3}
\end{equation}
These fix three components of $\omega^{~0j}_a$ represented by the
antisymmetric part of the matrix $M^{ij}_{}$, ~ 
$ M^{ij} -M^{ji}_{} ~=~0$,  leaving six components in the symmetric 
matrix $M^{ij}_{}$ undetermined. Lastly, Eqn.(\ref{eqomega1}) is solved by:
\begin{eqnarray}
&&\omega^{~ij}_a ~ =~ {\bar \omega}^{~ij}_a (e) ~+~ \kappa^{~ij}_a ~=~ 
{\bar \omega}^{~ij}_a (e) ~+~ \epsilon^{ijk}_{} N^{~k}_a ~,   
~~~ N^{~k}_a ~=~ e^l_a~N^{lk}_{} ~~ {\rm with} ~~
N^{lk}_{} ~=~ N^{kl}_{}~ , \nonumber \\
&&{\bar \omega}^{~ij}_a (e) ~\equiv~ \frac{1}{2}~ \left( {\hat e}^b_i \partial^{}_{[a} e^j_{b]} ~-~ {\hat e}^b_j \partial^{}_{[a} e^i_{b]} 
~-~ e^l_a ~{\hat e}^b_i {\hat e}^c_j \partial^{}_{[b} e^l_{c]}
\right) \label{solomega4}
\end{eqnarray}
Here ${\bar \omega}^{~ij}_a (e)$ is the torsion-free connection
satisfying
\begin{equation}
 D^{}_{[a} ({\bar \omega}) ~ e^i_{b]} ~=~0
 \end{equation}
 
Thus, finally, of all the twenty four components of the gauge 
fields $\omega^{~IJ}_\mu$, we have fixed twelve by equations 
of motion (\ref{eqomega1} - \ref{eqomega4}). Rest twelve represented 
by two $3\times3$ symmetric matrices $M^{ij}_{}$ and $N^{ij}_{}$ 
introduced in Eqns.(\ref{solomega3})  and  (\ref{solomega4})
are left undetermined.
 
Now we shall analyze the last Euler-Lagrange equation of motion
(\ref{eqtetrad}) for non-invertible tetrads. We start by listing   
various components of $T^{~\mu}_I$ of Eqn. (\ref{TS1}) for the 
degenerate tetrad (\ref{tetradD}):
\begin{eqnarray}
T^{~t}_0~ \equiv ~ {\hat e}~ {\hat T}^{~t}_0 ~&=&~{\hat e} 
\left[ {\hat e}^c_k ~\partial^{}_c \phi~ B^k_{} ~+~
\frac{1}{2}~ \left( B^{}_M B^M_{} ~-~ m^2_{} \phi^2_{}
\right) \right] \label{TSD1} \\
T^{~t}_i ~\equiv ~ {\hat e}~ {\hat T}^{~t}_i ~&=&~   
-~ {\hat e} ~{\hat e}^c_i ~\partial^{}_c \phi~ B^0_{}  \label{TSD2} \\
T^{~a}_0 ~\equiv ~ {\hat e}~ {\hat T}^{~a}_0 ~&=&~   
-~ {\hat e} ~{\hat e}^a_k ~\partial^{}_t \phi~ B^k_{}  \label{TSD3} \\
T^{~a}_i ~\equiv ~ {\hat e}~ {\hat T}^{~a}_i ~&=&~   
{\hat e} ~{\hat e}^a_i ~\partial^{}_t \phi~ B^0_{}  \label{TSD4} 
\end{eqnarray}
Since due to the matter equations of motion (\ref{eqBphi}) 
the scalar field $\phi$ does not have any $t$ dependence
the last two components vanish:
\begin{equation}
T^{~a}_0 ~=~ 0 ~, ~~~~~~ T^{~a}_i ~=~0 \label{TSD5}
\end{equation}
 
For degenerate tetrads (\ref{tetradD}), it is  convenient to split 
the sixteen equations of motion in (\ref{eqtetrad})  
into four sets of $1$, $3$, $3$ and $9$ equations as follows:
\begin{eqnarray}
\kappa^2_{}~ T^{~t}_0  ~\equiv~ {\hat e}~ \kappa^2_{}~ 
{\hat T}^{~t}_0 ~&=&~ -~ \frac{1}{2}~{\hat e} 
\left[ {\hat e}^b_k ~{\hat e}^c_l~ R^{~~kl}_{bc} (\omega) 
~-~ 2 \Lambda \right]  \label{eqtetradD1} \\
\kappa^2_{}~ T^{~t}_i  ~\equiv~ {\hat e}~ \kappa^2_{}~ 
{\hat T}^{~t}_i ~&=&~ {\hat e}~ {\hat e}^b_i~ {\hat e}^c_l 
~ R^{~~0l}_{bc} (\omega) \label{eqtetradD2}\\
\kappa^2_{}~ T^{~a}_0  ~\equiv~ {\hat e}~ \kappa^2_{}~ 
{\hat T}^{~a}_0 ~&=&~ {\hat e}~ {\hat e}^a_k~ {\hat e}^b_l 
~ R^{~~kl}_{tb} (\omega) \label{eqtetradD3} \\
\kappa^2_{}~ T^{~a}_i  ~\equiv~ {\hat e}~ \kappa^2_{}~ 
{\hat T}^{~a}_i ~&=&~ {\hat e}~ {\hat e}^a_{[i}~ {\hat e}^b_{j]} 
~ R^{~~0j}_{tb} (\omega) \label{eqtetradD4}
\end{eqnarray}
We now use Eqns.(\ref{TSD1}-\ref{TSD5}) in these equations. Using 
$ T^{~a}_0 ~=~0$ and $T^{~a}_i ~=~0$ in  the equations of motion
(\ref{eqtetradD3}) and (\ref{eqtetradD4}) respectively lead to:
\begin{eqnarray}
&&{\hat e}^a_i~ R^{~~ij}_{ta} (\omega) ~=~0 \label{solRS1}\\
&& R^{~~0i}_{ta} (\omega)~=~ \partial^{}_t M^{~i}_a ~=~0 
\label{solRS2}
\end{eqnarray}
where we have used $\omega^{~0l}_t ~=~0$ from Eqn.(\ref{solomega1})
and $\omega^{~0j}_a ~\equiv ~ M^{~j}_a$ from Eqn.(\ref{solomega3}).
Next, equation of motion (\ref{eqtetradD2}) and Eqn.(\ref{TSD2})
imply:
\begin{equation}
{\hat e}^b_l~ R^{~~0l}_{ab} (\omega)~=~ {\hat e}^b_l~ D^{}_{[a} 
({\bar \omega}) M^{~l}_{b]} ~=~ \kappa^2_{} ~ e^i_a ~
{\hat T}^{~t}_i ~=~ -~  \kappa^2_{}~ \partial^{}_a \phi~ B^0_{}
\label{solRS3}
\end{equation}
where we have used Eqn.(\ref{solomega3}) with $M^{ij} ~=~ M^{ji}$
and Eqn.(\ref{solomega4}) which implies $ {\hat e}^b_l ~
\kappa^{~lm}_b ~=~ 0$ due to $N^{ij}_{} ~=~N^{ji}_{}$.

Lastly we study the equation of motion (\ref{eqtetradD1}). For this,
breaking  $\omega^{~ij}_a $ into torsion-free part 
${\bar \omega}^{~ij}_a(e)$ and contorsion part as in 
Eqn.(\ref{solomega4}), we find that\footnote{Note that the sign
of $M^{~i}_{[a} M^{~j}_{b]}$ term is positive as against that in the 
Euclidean gravity studied earlier \cite{kaul-sandipan1}. This sign 
is due to the Lorentzian nature of the internal metric $\eta^{IJ}_{}$.}
\begin{eqnarray*}
R^{~~ij}_{ab} (\omega) ~&=&~ {\bar R}^{~~ij}_{ab} ({\bar \omega}) ~+ 
\epsilon^{ijk}_{} D^{}_{[a} ({\bar \omega}) N^{~k}_{b]}~
-~ N^{~i}_{[a} N^{~j}_{b]} ~+ ~ M^{~i}_{[a} M^{~j}_{b]}\\
&=&~ {\bar R}^{~~ij}_{ab} ({\bar \omega}) ~-~ 
\epsilon^{ijk}_{}~ e^l_{[a} ~D^{}_{b]} ({\bar \omega}) ~N^{lk}_{} ~+~ 
\left( M^{li}_{} M^{kj}_{} ~-~ N^{li}_{} N^{kj}_{} \right)
e^l_{[a}~ e^k_{b]}
\end{eqnarray*}
where ${\bar R}^{~~ij}_{ab} ({\bar \omega}) ~\equiv~ 
\partial^{}_{[a} {\bar \omega}^{~ij}_{b]} ~+~ 
{\bar \omega}^{~ik}_{[a} {\bar \omega}^{~kj}_{b]}$. Using this fact in 
the equation of motion (\ref{eqtetradD1}), we obtain the constraint:
\begin{eqnarray}
{\hat e}^b_k~ {\hat e}^c_l ~ {\bar R}^{~~kl}_{bc} (({\bar \omega}) ~-~ 
\left(M^{kl}_{} M^{lk}_{} - M^{kk}_{} M^{ll}_{} \right) ~+~
\left( N^{kl}_{} N^{lk}_{} - N^{kk}_{} N^{ll}_{} \right) ~-~
2~\Lambda  \nonumber \\
~=~ -~ 2 \kappa^2_{} ~ {\hat T}^{~t}_0 ~=~ -~ 2 \kappa^2_{}
\left[ {\hat e}^c_k ~\partial^{}_c \phi~ B^k_{} + \frac{1}{2} ~ 
\left( B^{}_M B^M_{} - m^2_{} \phi^2_{} \right) \right]
\label{solRS4}
\end{eqnarray}
where we have used the property that matrix $N^{ij}_{}$ is symmetric.
 
Thus, we have four new constraint equations in 
(\ref{solRS1}-\ref{solRS4}) in addition to those in 
(\ref{solomega1}-\ref{solomega4}) obtained earlier. Note that
constraint (\ref{solRS1}) does not give us any additional 
information as it is identically satisfied when constraints 
(\ref{solomega1}-\ref{solomega4}) are used. This can readily be
seen by noting that $R^{~~ij}_{ta} (\omega) ~= \partial^{}_t 
\omega^{~ij}_a ~=~ \partial^{}_t \kappa^{~ij}_a$ when constraints 
constraints (\ref{solomega1}), ({\ref{solomega2}) and 
(\ref{solomega4}) hold. Now ${\hat e}^a_i ~ R^{~~ij}_{ta} (\omega)
~=~ \partial^{}_t \left( {\hat e}^a_i~ \kappa^{~ij}_a \right) ~=~0$
because ${\hat e}^a_i ~\kappa^{~ij}_a ~\equiv~ \epsilon^{ijk}_{}~
{\hat e}^a_i N^{~k}_a ~=~ \epsilon^{ijk}_{} N^{ik}_{} ~\equiv ~0$ 
due to the symmetric character of the matrix $N^{ij}_{}$.

A particular solution of constraints (\ref{solRS2}) and (\ref{solRS3})
is provided by:
\begin{equation}
M^{~i}_a ~=~ \lambda ~e^i_a~, ~~~~~ \Leftrightarrow ~~~ 
M^{ij}_{} ~=~ \lambda ~\delta^{ij}_{} \label{solRS23}
\end{equation}
where 
\begin{equation}
\partial^{}_t \lambda~=~ 0~,  ~~~~~~ \partial^{}_a \lambda
~=~ \frac{\kappa^2_{}}{2}~e^i_a~{\hat T}^{~t}_i ~=~ 
-~ \frac{\kappa^2_{}}{2}~\partial^{}_a \phi~ B^0_{} \label{solRS23'}
\end{equation}
Using this in the constraint (\ref{solRS4}) leads to the master
constraint:
\begin{eqnarray}
{\hat e}^b_k~ {\hat e}^c_l~ {\bar R}^{~~kl}_{bc} ({\bar \omega}) 
+ 6 \lambda^2_{} - 2\Lambda - \xi ~
&=&~ -~ 2 \kappa^2_{}~ {\hat T}^{~t}_0  \nonumber\\
~&=&~ -~ 2 \kappa^2_{} \left[ {\hat e}^c_k ~\partial^{}_c \phi~ B^k_{}
+ \frac{1}{2} \left( B^{}_M B^M_{} - m^2_{} \phi^2_{} \right) \right]
\label{masterRS}
\end{eqnarray}
where
\begin{equation}
\xi~\equiv~ N^{kk}_{} N^{ll}_{} - N^{kl}_{} N^{lk}_{} \label{xi}
\end{equation}
Note from Eqn.(\ref{masterRS}), we have
\begin{equation}
\partial^{}_t \xi~=~ 2\kappa^2_{}~ \partial^{}_t {\hat T}^{~t}_0 ~=~ 
\kappa^2_{} \left[ {\hat e}^c_k ~\partial^{}_c \phi +
B^{}_k \right] \partial^{}_t B^k_{} \label{xiS}
\end{equation}
where we use the fact that, by equations of motion, all   
fields except $N^{ij}_{}$ (and hence $\xi$) and $B^{}_i$  are 
$t$ independent  in (\ref{masterRS}).

This completes our analysis of  all the Euler-Lagrange equations 
of motion for degenerate tetrad (\ref{tetradD}). The connection
fields $\omega^{~IJ}_\mu$ are all given by 
(\ref{solomega1}-\ref{solomega4}) and (\ref{solRS23}, \ref{solRS23'}).
Further, we have the master constraint (\ref{masterRS}) relating
geometric quantities to the matter fields.

Lastly, we analyze the equation (\ref{TS2}). For degenerate tetrads
(\ref{tetradD}), this equation is identically satisfied for $I~=~i$;
both the left-hand side and right-hand side   are zero.
For $I~=~0$, this equation is exactly the same as (\ref{xiS}).

\section{First order Abelian gauge field action}

Now we consider vector gauge fields coupled to gravity. The 
discussion will be developed in detail for $U(1)$ vector gauge 
field. Generalization to more general  non-Abelian vector 
gauge fields is straight forward.
 
For an  Abelian vector gauge field $A^{}_\mu$ coupled to gravity, 
we start with the action:

\begin{equation}
S~=~ S^{}_{HP\Lambda} ~+~ S^{}_{EM} \label{GV}
\end{equation}
where $S^{}_{HP\Lambda}$ is the  Hilbert-Palatini action 
functional with cosmological constant (\ref{HP}) and the   
matter action functional is
\begin{equation}
S^{}_{EM}~=~ \frac{1}{8} \int d^4{}x ~\epsilon^{\mu\nu\alpha\beta}_{}  
\epsilon^{}_{IJKL} ~ e^K_\alpha e^L_\beta \left[ F^{}_{\mu\nu} 
~B^{IJ}_{} ~+~ \frac{1}{12}~ e^I_\mu e^J_\nu ~ B^{}_{MN} 
B_{}^{MN} \right] \label{AV}
\end{equation}
Here $F^{}_{\mu\nu} ~\equiv~ \partial^{}_\mu A^{}_\nu 
~-~ \partial^{}_\nu A^{}_\mu $ is the field strength of the 
vector gauge field $A^{}_\mu$ and six additional fields are 
introduced through  $B_{}^{MN}$ which is antisymmetric in the
$SO(1,3)$ internal space labels: $B_{}^{MN} ~= ~ -~ B_{}^{NM}$. 
Like $S^{}_{HP\Lambda}$, the matter field action functional 
$S^{}_{EM}$ is first order and is defined for both non-degenerate 
and degenerate tetrads.

Varying the action functional (\ref{GV}) with respect to $B_{}^{MN}$ 
and $A^{}_\mu$, respectively leads to the following Euler-Lagrange
equations of motion:
\begin{eqnarray}
&&\epsilon^{\mu\nu\alpha\beta}_{}  \epsilon^{}_{IJKL} 
~ e^K_\alpha e^L_\beta \left[ F^{}_{\mu\nu}~\delta^I_M \delta^J_N 
~+~ \frac{1}{6} ~e^I_\mu e^J_\nu ~B^{}_{MN} \right] ~=~0 
\label{eqBMN}\\
&&\epsilon^{\mu\nu\alpha\beta}_{}  \epsilon^{}_{IJKL} ~\partial^{}_\nu
\left( e^K_\alpha e^L_\beta ~B^{IJ}_{} \right) ~=~0 \label{eqA}
\end{eqnarray}
Note that these equations of motion are defined for both invertible
and non-invertible tetrads. As we shall see below, for non-degenerate 
tetrads, these two first order equations are equivalent to 
standard Maxwell equations of motion for electromagnetic field 
in curved space time.

Now varying the action (\ref{GV}) with respect to the connection
$\omega^{~IJ}_\mu$ and tetrad $e^I_\mu$ respectively yields the
Euler-Lagrange equations of motion:
\begin{eqnarray}
&&\epsilon^{\mu\nu\alpha\beta}_{} ~\epsilon^{}_{IJKL}~e^K_\nu 
~D^{}_\alpha (\omega) e^L_\beta~ = ~0 \label{eqomegaV}\\
&&\epsilon^{\mu\nu\alpha\beta}_{} ~\epsilon^{}_{IJKL}~e^J_\nu 
\left[ R^{~~~KL}_{\alpha \beta} (\omega) ~-~ \frac{2\Lambda}{3}~ 
e^K_\alpha e^L_\beta \right] ~ = ~ - 4\kappa^2_{} ~T^{~\mu}_I
\label{eqtetradV}
\end{eqnarray}
which are same as the equations (\ref{eqomega}) and (\ref{eqtetrad}) 
obtained for the scalar matter theory except for $T^{~\mu}_I$ on 
the right side of the second equation which, for the Abelian vector 
field matter, is now given by:
\begin{equation}
T^{~\mu}_I ~\equiv~ \frac{1}{4}~ \epsilon^{\mu\nu\alpha\beta}_{}
 ~\epsilon^{}_{IJKL}~ e^J_\nu \left( F^{}_{\alpha \beta} ~
 B^{KL}_{} ~ + ~\frac{1}{6}~ e^K_\alpha e^L_\beta ~B^{}_{MN} B^{MN}_{} \right) \label{TV1}
 \end{equation}
Also from (\ref{eqtetradV}), using equation of motion (\ref{eqomegaV}) 
and Bianchi identity, we notice that this  $T^{~\mu}_I$ for the vector 
matter field  has to obey the following equation:
\begin{equation}
 4\kappa^2_{}~D^{}_\mu(\omega) T^{~\mu}_I ~=~ - 
 ~ \epsilon^{\mu\nu\alpha\beta}_{} ~\epsilon^{}_{IJKL}~ 
 D^{}_\mu(\omega) e^J_\nu  ~R^{~~~KL}_{\alpha \beta} (\omega)
 \label{TV2}
\end{equation}

Like the action functional (\ref{GV}) and  the matter equations of 
motion (\ref{eqBMN}) and (\ref{eqA}),  the gravity equations of motion
(\ref{eqomegaV}) and (\ref{eqtetradV}) are also defined for 
both invertible and non-invertible tetrads.

\subsection{Non-degenerate tetrads}

For invertible tetrads, $del~e^I_\mu ~\equiv~ e ~\ne~0$, we can 
use the identities: $\epsilon^{\mu\nu\alpha\beta}_{} 
~\epsilon^{}_{IJKL}~e^K_\alpha e^L_\beta ~$ $=~2e~ e^{[\mu}_I 
e^{\nu]}_J$ and $\epsilon^{\mu\nu\alpha\beta}_{} 
~\epsilon^{}_{IJKL}~e^K_\alpha e^L_\beta e^I_\mu e^J_\nu ~=~24e$
to rewrite the matter action functional (\ref{AV}) as:
\begin{equation}
S^{}_{EM}~=~ \frac{1}{2} \int d^4_{}x~ e \left( e^\mu_I e^\nu_J~
 F^{}_{\mu\nu}~ B^{IJ}_{} ~+~ \frac{1}{2}~ B^{}_{MN}B^{MN}_{} 
 \right) \label{AV1}
\end{equation}
and Euler-Lagrange equations of motion (\ref{eqBMN}) and 
(\ref{eqA}) respectively can be written as:
\begin{eqnarray}
 e^\mu_M e^\nu_N ~F^{}_{\mu\nu} ~+ ~ B^{}_{MN} ~=~0 \label{eqBMN1}\\
 \partial^{}_\nu \left( e ~e^\mu_I e^\nu_J ~ B^{IJ}_{} \right) ~=~ 0
 \label{eqA1}
\end{eqnarray}
The first equation reflect the fact that $B^{}_{MN}$ are not independent 
fields. Use this constraint equation in the second equation 
(\ref{eqA1}) to obtain the second order equation as:
\begin{equation}
 \partial^{}_\nu\left(e~ g^{\mu \alpha}_{} g^{\nu \beta}_{}
~ F^{}_{\alpha\beta}  \right) ~=~0 \label{eqA2}
\end{equation}
Substitute the  constraint (\ref{eqBMN1}) in the matter action 
(\ref{AV1}) to write it as:
\begin{equation}
 S^{}_{EM}~=~-~ \frac{1}{4} ~\int d^4_{} x~ e ~F^{}_{\mu \nu}~
 F^{}_{\alpha \beta}~ g^{\mu\alpha}_{} g^{\nu \beta}_{}  \label{AV2} 
\end{equation} 
which is the standard second order form of the action functional
with (\ref{eqA2}) as the equation of motion for electromagnetic 
field in curved spacetime.

For invertible tetrads, using identities
$ \epsilon^{\mu\nu\alpha\beta}_{} ~\epsilon^{}_{IJKL}~ e^J_\nu 
 ~=~ e~ e^\mu_{[I} e^\alpha_K e^\beta_{L]}$ and  
$ \epsilon^{\mu\nu\alpha\beta}_{} ~\epsilon^{}_{IJKL}~ e^J_\nu 
e^K_\alpha e^L_\beta $ $~=~ 6e e^\mu_I $, equation (\ref{TV1}}) 
can be written as:
\begin{eqnarray*}
T^{~\mu}_I~=~ \frac{e}{4} \left[ -~4 F^{}_{IK} B^{\nu K} ~+~
e^\nu_I \left( 2 F^{}_{KL} B^{KL}_{} 
~+~ B^{}_{KL} B^{KL}_{} \right) \right] \label{TV3}
\end{eqnarray*}
where $F^{}_{IK} ~\equiv ~e^\mu_I e^\nu_K ~F^{}_{\mu\nu}$  and 
$B^{\mu I}_{} ~\equiv~ e^\mu_K B^{KI}_{}$.
This, on using the constraint (\ref{eqBMN1}), becomes:
\begin{eqnarray}
T^{~\nu}_I ~\equiv~ e ~{\tilde T}^{~\nu}_I ~=~ 
e~\left[ F^{\nu K}_{} F^{}_{IK} ~ -~ \frac{1}{4}
~ e^\nu_I ~F^{}_{\alpha \beta} F^{\alpha \beta}_{} \right]
\label{TV4}
\end{eqnarray}
 
The Euler-Lagrange equation of motion (\ref{eqomegaV}) is the same 
as that in the case of scalar matter field discussed earlier and,
for invertible tetrads, is solved exactly by the  torsion-free 
connection fields given in terms of the tetrads, $\omega^{~IJ}_\mu 
= \omega^{~IJ}_\mu (e)$ as in Eqn.(\ref{omega(e)}). Also for this
torsion-free connection, from Eqn.(\ref{TV2}) we notice that
$T^{~\mu}_I$ of Eqn.(\ref{TV4}) has to satisfy the condition:
\begin{equation}
 D^{}_\mu(\omega) T^{~\mu}_I \equiv~  D^{}_\mu ( \omega) 
\left( e~{\tilde T}^{~\mu}_I\right)~=~  0
 \label{TV5}
 \end{equation}

Same discussion as was done for the scalar matter field case 
earlier, allows us to write the other Euler-Lagrange equation 
(\ref{eqtetradV}) for invertible tetrads as:
\begin{equation} \label{eqtetradV1}
 R^{~\mu}_I ~-~ \frac{1}{2}~ e^\mu_I ~R ~=~ \kappa^2_{}~ 
 {\tilde T}^{~\mu}_I ~- e^\mu_I ~ \Lambda
\end{equation}
where $R^{~\mu}_I ~\equiv~ e^\alpha_I e^\beta_L e^\mu_K ~ 
R^{~~~KL}_{\alpha\beta}(\omega)$ and $R ~\equiv~ e^I_\mu ~ 
R^{~\mu}_I~\equiv~ e^\alpha_I e^\beta_J ~R^{~~~IJ}_{\alpha\beta} 
(\omega)$  with  now ${\tilde T}^{~\mu}_I$ for electro-magnetic 
field  given by of Eqn.(\ref{TV4}).  Clearly, Eqns.(\ref{eqtetradV1})
are the standard second order Einstein field equations for gravity 
with electromagnetic matter obtained in the usual second order  
formalism. Note that ${\tilde T}^{\mu\nu} = e^{\mu_ I}_{} ~
{\tilde T}^{~\nu}_I = \left( F^{\mu\alpha}_{} F^{\nu}_{~~\alpha}  
-  \frac{1}{4}~g^{\mu\nu} ~F^{}_{\alpha \beta} 
F^{\alpha \beta}_{} \right) $ is the standard stress-energy tensor 
for the electromagnetic field and  Eqn.(\ref{TV5}) represents 
the conservation condition for this stress-energy  tensor. 
 
\subsection{Degenerate tetrads}

Now we analyze the general Euler-Lagrange equations of motion
(\ref{eqBMN} - \ref{eqtetradV}) for tetrads with one zero 
eigenvalue as in (\ref{tetradD}). Matter equations of motion 
(\ref{eqBMN}) and (\ref{eqA}) for this degenerate tetrad imply:
\begin{equation}
 F^{}_{ta}~=~0~; ~~~~~~\partial^{}_t \left( {\hat e}~ {\hat e}^a_i
 ~ B^{0i}_{} \right) ~=~0 ~, ~~~~~ \partial^{}_a \left( 
 {\hat e} ~{\hat e}^a_i~ B^{0i}_{} \right) ~=~0 \label{eqBA}
\end{equation}
We have no conditions on $F^{}_{ab}$ and $B^{}_{ij}$.

Euler-Lagrange equation of motion (\ref{eqomegaV}) here is the
same as that for the scalar case, and hence the connection 
components here are the same:
\begin{eqnarray}
&&\partial^{}_t e^i_a ~=~0~;~~~~ \omega^{~0i}_t ~=~0,
 ~~ \omega^{~ij}_t ~=~0~;  \nonumber \\
&&\omega^{~oj}_a ~\equiv ~M^{~j}_a \equiv e^i_a ~M^{ij}_{} 
 ~~{\rm with} ~~M^{ij}_{} ~=~M^{ji}_{}~; \nonumber \\
&&\omega^{~ij}_a ~=~ {\bar \omega}^{~ij}_a (e) ~
 + ~ \kappa^{~ij}_a ~\equiv~ {\bar \omega}^{~ij}_a (e) 
 ~+~ \epsilon^{ijk}_{} N^{~k}_a~, \nonumber\\
&&  ~~~~~~~~~ ~~N^{~k}_a ~=~ e^l_a N^{lk}_{} 
 ~ ~{\rm with }~~~ N^{lk}_{} ~=~ N^{kl}_{}~, \nonumber \\
&& ~~~~~~~~~~~{\bar \omega}^{~ij}_a (e) ~\equiv~ \frac{1}{2}~ 
\left( {\hat e}^b_i \partial^{}_{[a} e^j_{b]} ~-~ {\hat e}^b_j
\partial^{}_{[a} e^i_{b]} 
~-~ e^l_a ~{\hat e}^b_i {\hat e}^c_j \partial^{}_{[b} e^l_{c]}
\right)
 \label{omegaV}
\end{eqnarray}
where we have made a gauge choice to make the triads $e^i_a$ as
$t$-independent. Use these solutions to write the matter 
Euler-Lagrange equations of motion (\ref{eqBA}) as:
\begin{eqnarray}
&& F_{ta}^{} ~=~0~; ~~~~~~~\partial^{}_t B^{0i}_{}~=~0~, \nonumber \\
&& {\hat e}^a_iD^{}_a ({\bar \omega}) B^{0i}_{} ~\equiv ~ 
{\hat e}^a_i \left(\partial^{}_a B^{0i}_{} + 
{\bar \omega}^{~ij}_a B^{0j}_{} \right)~
=~{\hat e}^a_i \left( \partial^{}_a  B^{0i}_{} 
+ {\omega}^{~ij}_a B^{0j}_{} \right)~=~0
\end{eqnarray}
where we have used $D^{}_a (\omega)\left({\hat e} 
{\hat e}^a_i \right) ~=~ D^{}_a ({\bar \omega}) 
\left({\hat e} {\hat e}^a_i \right) ~=~0$ to obtain  the last equation.

The various components of $T^{~\mu}_I$ of (\ref{TV1}) for
degenerate tetrad (\ref{tetradD}) now can been written as:
\begin{eqnarray}
T^{~t}_0 ~\equiv ~ {\hat e}~ {\hat T}^{~t}_0 ~&=&~ \frac{{\hat e}}{2}
\left( {\hat e}^b_k {\hat e}^c_l ~ B^{kl}_{} ~ F^{}_{bc}
~+~ \frac{1}{2} ~B^{}_{MN} B^{MN}_{} \right) \nonumber \\
T^{~t}_i ~\equiv~ {\hat e}~ {\hat T}^{~t}_i 
~&=&~ {\hat e}~ {\hat e}^b_l {\hat e}^c_i~ 
F^{}_{bc} ~B^{0l}_{} \nonumber \\
~T^{~a}_0 ~\equiv~ {\hat e}~ {\hat T}^{~a}_0 
~&=&~ {\hat e}~ {\hat e}^a_k {\hat e}^b_l~ 
F^{}_{bt} ~B^{kl}_{} \nonumber \\
~T^{~a}_i ~\equiv~ {\hat e}~ {\hat T}^{~a}_i 
~&=&~ -~ {\hat e}~ {\hat e}^a_{[i} {\hat e}^b_{l]}~ 
F^{}_{bt} ~B^{0l}_{}  \label{TVD1}
\end{eqnarray}
Since $F^{}_{at} ~=~0$ due to the matter equation of motion above,
the last two equations here are:
\begin{equation}
T^{~a}_0 ~\equiv ~ {\hat e}~ {\hat T}^{~a}_0 ~=~ 0~, ~~~~~~~
T^{~a}_i ~\equiv ~ {\hat e}~ {\hat T}^{~a}_i ~=~ 0 \label{TVD2}
\end{equation}
 
Like in the scalar matter theory, the sixteen Euler-Lagrange 
equations of motion in (\ref{eqtetradV})  for degenerate tetrads
are:
\begin{eqnarray}
\kappa^2_{}~ T^{~t}_0  ~\equiv~ {\hat e}~ \kappa^2_{}~ 
{\hat T}^{~t}_0 ~&=&~ -~ \frac{1}{2}~{\hat e} 
\left[ {\hat e}^b_k ~{\hat e}^c_l~ R^{~~kl}_{bc} (\omega) 
~-~ 2 \Lambda \right]  \label{eqtetradDV1} \\
\kappa^2_{}~ T^{~t}_i  ~\equiv~ {\hat e}~ \kappa^2_{}~ 
{\hat T}^{~t}_i ~&=&~ {\hat e}~ {\hat e}^b_i~ {\hat e}^c_l 
~ R^{~~0l}_{bc} (\omega) \label{eqtetradDV2}\\
\kappa^2_{}~ T^{~a}_0  ~\equiv~ {\hat e}~ \kappa^2_{}~ 
{\hat T}^{~a}_0 ~&=&~ {\hat e}~ {\hat e}^a_k~ {\hat e}^b_l 
~ R^{~~kl}_{tb} (\omega) \label{eqtetradDV3} \\
\kappa^2_{}~ T^{~a}_i  ~\equiv~ {\hat e}~ \kappa^2_{}~ 
{\hat T}^{~a}_i ~&=&~ {\hat e}~ {\hat e}^a_{[i}~ {\hat e}^b_{j]} 
~ R^{~~0j}_{tb} (\omega) \label{eqtetradDV4}
\end{eqnarray}
but with $T^{~\mu}_I$ now for vector gauge matter fields 
as in (\ref{TVD1}). As earlier, Eqns.(\ref{eqtetradDV3}) and
(\ref{eqtetradDV4}) respectively imply:
\begin{eqnarray}
&&{\hat e}^a_i~ R^{~~ij}_{ta} (\omega) ~=~0 \label{solRV1}\\
&& R^{~~0i}_{ta} (\omega)~=~ \partial^{}_t M^{~i}_a ~=~0 
\label{solRV2}
\end{eqnarray}
where we have used (\ref{TVD2}) and (\ref{omegaV}). Next, use 
the second equation in (\ref{TVD1}) in the equation of motion
(\ref{eqtetradDV2}), to obtain the constraint:
\begin{equation}
{\hat e}^b_l ~R^{~~0l}_{ab} (\omega) ~=~ {\hat e}^b_l~ 
D^{}_{[a} ({\bar \omega}) M^{~l}_{b]} ~=~ \kappa^2_{} ~e^i_a ~
{\hat T}^{~t}_i ~= ~ - ~ \kappa^2_{}~ {\hat e}^b_l ~F^{}_{ab} 
~B^{0l}_{} 
\label{solRV3}
\end{equation}
Lastly, using the first equation in (\ref{TVD1}) in the 
equation of motion (\ref{eqtetradDV1}), we have 
\begin{eqnarray*}
 {\hat e}^b_k {\hat e}^c_l~ R^{~~kl}_{bc} (\omega ) ~- 
 ~ 2\Lambda ~=~  -2 \kappa^2_{}~ {\hat T}^{~t}_0 ~
 =~ -~ \kappa^2_{} \left( {\hat e}^b_k {\hat e}^c_l~ F^{}_{bc}~
  B^{kl}_{} + \frac{1}{2}~ B^{}_{MN} B^{MN}_{} \right)  
\end{eqnarray*}
which, using $ \omega ^{~ij}_a ~=~ {\bar \omega}^{~ij}_a(e)
+\epsilon^{ijk}_{} N^{~k}_a$, can be further seen to be:
\begin{eqnarray}
 {\hat e}^b_k {\hat e}^c_l~ {\bar R} ^{~~kl}_{bc} ({\bar \omega} ) 
 ~-~ \left( M^{kl}_{} M^{lk}_{} - M^{kk}_{} M^{ll}_{} \right)
 ~+~ \left( N^{kl}_{} N^{lk}_{} - N^{kk}_{} N^{ll}_{} \right)
 ~-~ 2 \Lambda  \nonumber \\
 ~=~ -~ 2\kappa^2_{} ~{\hat T}^{~t}_0 ~=~ -~ 
 \kappa^2_{} \left( {\hat e}^b_k {\hat e}^c_l~ F^{}_{bc}~
  B^{kl}_{} + \frac{1}{2}~ B^{}_{MN} B^{MN}_{} \right) 
  \label{solRV4}
\end{eqnarray}

Note that (\ref{solRV1}) holds identically for configurations
satisfying the constraints (\ref{omegaV}) and a particular
solution to the constraints (\ref{solRV2}) and (\ref{solRV3}) is
given by
\begin{eqnarray}
M^{~i}_a ~=~ \lambda ~ e^i_a~, ~~~\partial^{}_t \lambda ~=~0~,
~~~~ \partial^{}_a \lambda ~=~ \frac{\kappa^2_{}}{2}~ e^i_a 
{\hat T}^{~t}_i ~=~ -~ \frac{\kappa^2_{}}{2} ~ {\hat e}^b_l ~F^{}_{ab} 
~B^{0l}_{} \label{solRV5}
\end{eqnarray}
Using this in Eqn.(\ref{solRV4}), we have the final master constraint:
\begin{eqnarray}
{\hat e}^b_k {\hat e}^c_l~ {\bar R} ^{~~kl}_{bc} ({\bar \omega} ) 
~+~ 6\lambda^2_{} ~-~ 2 \Lambda  ~-~ \xi  ~&=& ~ 
-~ 2\kappa^2_{} ~{\hat T}^{~t}_0 ~ \nonumber \\
~&=&~ -~ \kappa^2_{} \left( {\hat e}^b_k {\hat e}^c_l~ F^{}_{bc}~
B^{kl}_{} + \frac{1}{2}~ B^{}_{MN} B^{MN}_{} \right) 
\label{masterconstV}
\end{eqnarray}
where $\xi ~\equiv~ N^{kk}_{} N^{ll}_{} - N^{kl}_{} N^{lk}_{} $.

Note that, from the constraint (\ref{masterconstV}), using  
$\partial^{}_t F^{}_{bc} \equiv -\left( \partial^{}_b F^{}_{ct} 
+ \partial^{}_c F^{}_{tb} \right) = 0$ due to matter equation 
of motion, we have:
\begin{equation}
\partial{}_t \xi~=~  2\kappa^2_{} ~\partial^{}_t {\hat T}^{~t}_0 
~=~ \kappa^2_{} \left( {\hat e}^b_k {\hat e}^c_l ~F^{}_{bc} 
+ B^{}_{kl} \right) \partial^{}_t B^{kl}_{} \label{solRV6}
\end{equation}
This equation is equivalent to Eqn.(\ref{TV2}) for $I=0$ for
degenerate tetrads (\ref{tetradD}). For $I=i$, equation
(\ref{TV2}) is identically satisfied, as both left-hand
and right-hand side are zero   for degenerate tetrads.

\section{Fermion action functional}
Here, for a fermion coupled to gravity in the first order formulation,
we start with the following action:
\begin{equation}
 S~=~ S^{}_{HP\Lambda} ~+ ~ S^{}_F \label{AGF}
\end{equation}
where $S^{}_{HP \Lambda}$ is the Hilbert-Palatini action (\ref{HP})
and the fermion matter action is{\footnote{Our gamma 
matrices satisfying the Clifford algebra 
$\gamma^I_{} \gamma^J_{} + \gamma^J_{} \gamma^I_{} = - 2 \eta^{IJ}_{}$,
with $\eta^{IJ}_{}$ $ =  \eta^{}_{IJ}$ $ =  dia\left(-1,+1,+1,+1\right)$,
are $ \gamma^0{}=\left(\begin{array}{cc}0&{\mathbf 1}
\\{\mathbf 1}&0\end{array}\right) $, $ \gamma^i{}=\left(\begin{array}{cc}
0&\sigma^i_{}\\-\sigma^i_{}&0\end{array}\right) $, with $\gamma^{}_5 =
-i \gamma^0{} \gamma^1_{} \gamma^2_{} \gamma^3_{} 
= i \gamma^{}_0 \gamma^{}_1\gamma^{}_2 \gamma^{}_3 = 
\left(\begin{array}{cc}{\mathbf 1}&0\\ 0&-{\mathbf 1}  
\end{array}\right)$.
Note that $\left( \gamma^0{} \right)^\dagger_{} =
\gamma^0_{}$ ,  $ \left( \gamma^i_{}\right)^\dagger_{} = 
- \gamma^i_{}$; $~\left( \sigma^{0i}_{} \right)^\dagger
 = \sigma^{0i}_{} $,
$\left( \sigma^{ij}_{} \right)^\dagger = -\sigma^{ij}_{} $
and $\sigma^{}_{LK} \gamma^I{} - \gamma^I_{} \sigma^{}_{LK} 
= \delta^I_{[L} \gamma^{}_{K]}$; $ ~ \gamma^I_{} \sigma_{}^{LK} 
+\sigma_{}^{LK} \gamma^I{} =i \epsilon^{ILKM}_{} 
\gamma^{}_5 \gamma^{}_M$.}:
\begin{eqnarray}
S^{}_F ~=~ \frac{1}{6} \int d^4_{}x~ \epsilon^{\mu\nu\alpha \beta}_{}
\epsilon^{}_{IJKL} ~e^J_\nu e^K_\alpha e^L_\beta \left[ \frac{i}{2}
{\overline \psi} ~\gamma^I_{} D^{}_\mu (\omega) \psi -\frac{i}{2}
{\overline {D^{}_\mu( \omega) \psi}} ~\gamma^I_{}   \psi  + 
\frac{m}{4}  ~e^I_\mu ~{\overline \psi} \psi \right] \label{AF}
\end{eqnarray}
where $SO(1,3)$ covariant derivatives are $D^{}_\mu(\omega) 
\psi\equiv \partial^{}_\mu \psi- \frac{1}{2}
\omega^{~IJ}_\mu \sigma^{}_{IJ} \psi$ and 
${\overline {D^{}_\mu(\omega) \psi}}\equiv \left( D^{}_\mu 
(\omega) \psi \right)^\dagger_{} \gamma^0_{}\equiv 
\partial^{}_\mu {\overline \psi}+ \frac{1}{2}
\omega^{~IJ}_\mu {\overline \psi} ~\sigma^{}_{IJ} $  with
$\sigma^{}_{IJ} = \frac{1}{4} [ \gamma^I_{}, ~ 
\gamma^J_{}]$. Note that fermion action $S^{}_F$, like
Hilbert-Palatini action (\ref{HP}), is defined for 
both invertible and non-invertible tetrads.

Varying the total action (\ref{AGF}) with respect to the independent 
fields, ${\overline \psi}$, $\psi$, connection $\omega^{~IJ}_\mu$
and tetrad $e^I_\mu$, respectively leads to the Euler-Lagrange 
equations of motion: 
\begin{eqnarray}
&&\epsilon^{\mu\nu\alpha \beta}_{} \epsilon^{}_{IJKL}
~e^J_\nu e^K_\alpha e^L_\beta \left[ i \gamma^I_{} D^{}_\mu (\omega)
\psi + \frac{m}{4}~ e^I_\mu \psi \right] + \frac{3i}{2} ~
\epsilon^{\mu\nu\alpha \beta}_{} \epsilon^{}_{IJKL} ~S^{~~J}_{\mu\nu}
e^K_\alpha e^L_\beta \gamma^I_{} \psi ~=~0 \label{eqpsibar}\\
&&\epsilon^{\mu\nu\alpha \beta}_{} \epsilon^{}_{IJKL}
~e^J_\nu e^K_\alpha e^L_\beta \left[ - i {\overline { D^{}_\mu (\omega)
\psi}} ~\gamma^I_{} + \frac{m}{4}~ e^I_\mu {\overline \psi} \right] 
- \frac{3i}{2} ~ \epsilon^{\mu\nu\alpha \beta}_{} \epsilon^{}_{IJKL}
~S^{~~J}_{\mu\nu} e^K_\alpha e^L_\beta
{\overline \psi} \gamma^I_{}~=~0 \label{eqpsi}\\
&&\epsilon^{\mu\nu\alpha \beta}_{} \left[ \epsilon^{}_{IJKL}
e^K_\nu D^{}_\alpha (\omega) e^L_\beta ~+~ \frac{\kappa^2_{}}{2}
e^{}_{\nu I} e^{}_{\alpha J} e^{}_{\beta M} {\overline \psi}
\gamma^{}_5 \gamma^M_{} \psi \right] ~=~0  \label{eqomegaF}\\
&&\epsilon^{\mu\nu\alpha\beta}_{} ~\epsilon^{}_{IJKL}~e^J_\nu 
\left[ R^{~~~KL}_{\alpha \beta} (\omega) ~-~ \frac{2\Lambda}{3}~ 
e^K_\alpha e^L_\beta \right] ~=~ - 4\kappa^2_{} ~T^{~\mu}_I
\label{eqtetradF}
\end{eqnarray}
where $2 S^{~~I}_{\mu\nu} \equiv D^{}_{[\mu} (\omega) e^I_{\nu]}$ 
is the torsion and 
\begin{equation} 
T^{~\mu}_I ~=~ \frac{1}{4}~ \epsilon^{\mu\nu\alpha \beta}_{} 
\epsilon^{}_{IJKL}~ e^J_\nu e^K_\alpha \left[ i {\overline \psi} 
\gamma^L_{} D^{}_\beta (\omega) \psi - i {\overline {D^{}_\beta
( \omega) \psi}} ~\gamma^L_{}   \psi  + 
\frac{2m}{3} ~ e^L_\beta ~{\overline \psi} \psi \right] \label{TF1}
\end{equation}
Apply  covariant derivative   to Eqn.(\ref{eqtetradF}) 
and use Bianchi identity and Eqn.(\ref{eqomegaF}) to obtain 
the  constraint on this $T^{~\mu}_I$  as:
\begin{equation}
4\kappa^2_{}~D^{}_\mu(\omega) T^{~\mu}_I ~=~ - 
~ \epsilon^{\mu\nu\alpha\beta}_{} ~\epsilon^{}_{IJKL}~ 
D^{}_\mu(\omega) e^J_\nu  ~R^{~~~KL}_{\alpha \beta} (\omega)
\label{TF2}
\end{equation}

Like in the earlier cases, all the Euler-Lagrange equations of 
motion above are defined for both non-degenerate and degenerate 
tetrads.
 
\subsection{Invertible tetrads}
For non-degenerate tetrads, it is straight forward to check that 
the fermion action (\ref{AF}) can be written in the standard form:
\begin{equation}
S^{}_F ~=~ \int d^4_{}x ~ e~ \left[ \frac{i}{2}~ {\overline \psi} 
e^\mu_I \gamma^I_{} D^{}_\mu (\omega) \psi  ~- ~\frac{i}{2}~
{\overline {D^{}_\mu (\omega) \psi}}~ e^\mu_I \gamma^I_{} \psi 
~+~ m {\overline \psi} \psi \right] \label{AF1}
\end{equation}
and the fermion equations of motion (\ref{eqpsibar}) and (\ref{eqpsi}) 
take the form:
\begin{eqnarray}
i e^\mu_I \gamma^I_{} D^{}_\mu (\omega) \psi ~+~ m\psi ~
-~i e^\mu_I e^\nu_J S^{~~I}_{\mu\nu} \gamma^J_{} \psi ~=~0 
\nonumber\\
-~i e^\mu_I {\overline {D^{}_\mu (\omega) \psi}} ~\gamma^I_{}
+~ m{\overline \psi}  ~+~i e^\mu_I e^\nu_J S^{~~I}_{\mu\nu}
{\overline \psi} \gamma^J_{}  ~=~0 \label{eqF}
\end{eqnarray} 

Next, for invertible tetrads, the equation of motion (\ref{eqomegaF}) 
can be simplified to
\begin{equation}
2S^{}_{\alpha \beta I}~\equiv~ 
D^{}_{[\alpha} (\omega) e^{}_{\beta ] I}~=~ 
-~ \frac{\kappa^2_{}}{2}~\epsilon^{}_{IJKL} e^J_\alpha e^K_\beta 
~{\overline \psi} \gamma^{}_5 \gamma^L_{} \psi \label{torsion}
\end{equation}
indicating presence of matter induced torsion. This equation  
can be solved for the twenty four connection components as:
\begin{equation}
\omega^{~IJ}_\mu ~=~ \omega^{~IJ}_\mu (e) ~+~ \kappa^{~IJ}_\mu 
\label{omegaFer}
\end{equation}
where $\omega^{~IJ}_\mu (e)$ is the torsion-free connection 
(\ref{omega(e)}), $
D^{}_{[\mu} (\omega (e) )e^I_{\nu]} ~=~ 0  $, and contorsion is:
\begin{equation}
 \kappa^{~IJ}_\mu~=~ \frac{\kappa^2_{}}{4} ~\epsilon^{IJKL}_{}~
 e^{}_{\mu K}~ {\overline \psi} \gamma^{}_5 \gamma^{}_L
  \psi \label{contorsion}
\end{equation}
Finally, equation of motion (\ref{eqtetradF}) for invertible tetrads
can be cast in the standard form:
\begin{equation} \label{eqtetradF1}
 R^{~\mu}_I ~-~ \frac{1}{2}~ e^\mu_I ~R ~=~ \kappa^2_{}~ 
 {\tilde T}^{~\mu}_I ~- e^\mu_I ~ \Lambda
\end{equation}
where $R^{~\mu}_I ~\equiv~ e^\alpha_I e^\beta_L e^\mu_K ~ 
R^{~~~KL}_{\alpha\beta}(\omega)$ and $R ~\equiv~ e^I_\mu ~ 
R^{~\mu}_I~\equiv~ e^\alpha_I e^\beta_J ~R^{~~~IJ}_{\alpha\beta} 
(\omega)$  with  now ${\tilde T}^{~\mu}_I$  of ({\ref{TF1}) 
for fermion fields as:  
\begin{eqnarray}
T^{~\mu}_I~ ~\equiv ~ e~ {\tilde T}^{~\mu}_I ~=~ e~ \left[ e^\mu_I
\left( \frac{i}{2} ~{\overline \psi} ~e^\beta_L \gamma^L_{} 
~D^{}_\beta (\omega) \psi  - \frac{i}{2}~ {\overline { D^{}_\beta 
(\omega) \psi }} ~e^\beta_L \gamma^L_{}~ \psi + m {\overline \psi} 
~\psi \right) \right. \nonumber \\ 
\left. ~-~ e^\mu_L e^\beta_I \left( \frac{i}{2} ~{\overline \psi}
~\gamma^L_{} ~D^{}_\beta (\omega) \psi - \frac{i}{2}~
{\overline {D^{}_\beta (\omega) \psi}} ~\gamma^L_{}~ \psi \right) 
\right] \label{TF3}
\end{eqnarray}

Eqns.(\ref{eqtetradF1}) with the connection fields given 
by (\ref{omegaFer}) and (\ref{contorsion}), are the  second 
order  Einstein field  equations. However, notice that this 
theory is not exactly same as  the
standard second order  theory obtained from Einstein-Hilbert 
action with fermion matter described by an action obtained by 
minimal coupling prescription. This well known difference 
lies in Eqns.(\ref{torsion} - \ref{contorsion}) reflecting 
presence of fermion dependent torsion in the theory discussed 
above. This is in contrast to the standard second order theory 
of gravity with fermions which is torsion free.

\subsection{Non-invertible tetrads}
We now analyze the fermion theory for tetrads (\ref{tetradD})
with one zero eigen value. The fermion Euler-Lagrange equations of 
motion (\ref{eqpsibar}) and ({\ref{eqpsi}) for this degenerate
tetrad can be written as:
\begin{eqnarray}
D^{}_t (\omega) \psi ~+~ {\hat e}^a_i~ \gamma^0{} 
\left( S^{~~i}_{ta} ~\gamma^0_{} - S^{~~0}_{ta} ~\gamma^i_{} \right) 
\psi ~=~0 \nonumber \\
{\overline {D^{}_t (\omega) \psi}} ~+~ {\hat e}^a_i~{\overline \psi}
\left( S^{~~i}_{ta} ~\gamma^0_{} - S^{~~0}_{ta} ~\gamma^i_{} \right) 
\gamma^0_{} ~=~0 \label{eqF1}
\end{eqnarray}

We break twenty four Euler-Lagrange equations of motion in 
(\ref{eqomegaF}) into four sets of $3$, $3$, $9$ and $9$ equations 
as:
\begin{eqnarray}
&&\epsilon^{abc}_{}~ \epsilon^{}_{jkl}~e^k_a D^{}_b(\omega) e^l_c ~=~~0 
\label{eqomegaF1}\\
&&\epsilon^{abc}_{}\left(\epsilon^{}_{ijk}~e^k_a D^{}_b(\omega) e^0_c
~-~ \frac{\kappa^2_{}}{2}~ e^{}_{ai} e^{}_{bj}e^{}_{ck} 
{\overline \psi}  \gamma^{}_5 \gamma^k_{} \psi\right) ~=~~0 
\label{eqomegaF2}\\
&&\epsilon^{abc}_{}~ \epsilon^{}_{ijk}~e^j_b D^{}_t(\omega) e^k_c ~=~~0 
\label{eqomegaF3}\\
&&\epsilon^{abc}_{}~ \epsilon^{}_{ijk}~e^j_b D^{}_t(\omega) e^0_c ~=~~0
\label{eqomegaF4}  
\end{eqnarray}
Except for (\ref{eqomegaF2}), all these equations are the same as those
obtained for the scalar and vector theories above. So we can follow 
similar discussion as earlier to solve these. The solution of 
equations (\ref{eqomegaF1}), (\ref{eqomegaF3}) and (\ref{eqomegaF4}) 
is given by:
\begin{eqnarray}
&&\partial^{}_t e^i_a ~=~0~;~~~~ \omega^{~0i}_t ~=~0,
~~ \omega^{~ij}_t ~=~0~;  \nonumber \\
&&\omega^{~ij}_a ~=~ {\bar \omega}^{~ij}_a (e) ~
+ ~ \kappa^{~ij}_a ~\equiv~ {\bar \omega}^{~ij}_a (e) 
 ~+~ \epsilon^{ijk}_{} N^{~k}_a~, \nonumber\\
&&  ~~~~~~~~~ ~~N^{~k}_a ~=~ e^l_a ~N^{lk}_{} 
~ ~{\rm with }~~~ N^{lk}_{} ~=~ N^{kl}_{}~, \nonumber \\
&& ~~~~~~~~~~~~{\bar \omega}^{~ij}_a (e) ~\equiv~ \frac{1}{2}~ 
\left( {\hat e}^b_i \partial^{}_{[a} e^j_{b]} ~-~ {\hat e}^b_j 
\partial^{}_{[a} e^i_{b]} ~-~ e^l_a ~{\hat e}^b_i {\hat e}^c_j 
\partial^{}_{[b} e^l_{c]}\right) \label{omegaF1}
\end{eqnarray}
where, without loss of generality, we have made the gauge choice 
to make the triads $e^i_a$ independent of $t$. The only change 
appears in the connection component $\omega^{~0i}_a$ obtained 
from (\ref{eqomegaF2}) which is now:
\begin{eqnarray}
&&\omega^{~0j}_a ~\equiv~  M^{~j}_a ~= ~e^i_a ~M^{ij}_{} ~= e^i_a 
~\left({\bar M}^{ij}_{} + {\tilde M}^{ij}_{} \right)~\equiv~ 
e^i_a \left( {\bar M}^{~j}_a +{\tilde M}^{~j}_a \right), \nonumber \\
&&{\bar M}^{ij}_{} ~\equiv ~\frac{1}{2}~ \left( M^{ij}_{} 
+M^{ji}_{} \right), ~~~ {\tilde M}^{ij}_{}~ \equiv~ 
\frac{1}{2}~\left( M^{ij}_{} - M^{ji}_{}\right) ~
=~ \frac{\kappa^2_{}}{2}~ \epsilon^{ijk}_{}~ 
{\overline \psi} \gamma^{}_5\gamma^{}_k \psi \label{omegaF2}
\end{eqnarray}
Thus, three components of $M^{ij}_{}$ represented by the antisymmetric
part ${\tilde M}^{ij}_{}$ are fixed in terms of the fermions, but other
six components in the symmetric part ${\bar M}^{ij}_{}$ are not
determined by the equations of motion. These are in addition to the
six undetermined fields contained in the symmetric matrix $N^{ij}_{}$
of (\ref{omegaF1}). This solves all the equations of motion 
in (\ref{eqomegaF1}-\ref{eqomegaF4}).

Note that Eqn.(\ref{omegaF1}), implies $ 2 S^{~~I}_{ta} \equiv
D^{}_t(\omega ) e^I_a - D^{}_a(\omega)e^I_t = D^{}_t (\omega) 
e^I_a =0$. This, in turn, for the fermion equations of motion
(\ref{eqF1}) implies:
\begin{eqnarray}
D^{}_t (\omega) \psi ~ =  ~ \partial^{}_t \psi ~=~0 \label{eqF2}
\end{eqnarray}

Various components of $T^{~\mu}_I$ of (\ref{TF1}) for degenerate
tetrads (\ref{tetradD}) can be written as:
\begin{eqnarray}
&&T^{~t}_0 ~\equiv~ {\hat e}~ {\hat T}^{~t}_0 ~= ~ {\hat e}
\left[ \frac{i}{2}~ {\hat e}^a_l \left(~ {\overline \psi} \gamma^l_{}
D^{}_a (\omega) \psi  - {\overline {D^{}_a (\omega) \psi}}
~ \gamma^l_{} \psi \right) ~+~ m {\overline \psi }\psi \right] 
\label{TFD1}\\
&&T^{~t}_i ~\equiv~ {\hat e}~ {\hat T}^{~t}_i ~= ~ -~\frac{i}{2}~ 
{\hat e} {\hat e}^a_i \left( ~ {\overline \psi} \gamma^0_{}
D^{}_a (\omega) \psi  - {\overline {D^{}_a (\omega) \psi}}
~ \gamma^0_{} \psi \right) \label{TFD2}\\
&&T^{~a}_0 ~\equiv~ {\hat e}~ {\hat T}^{~a}_0 ~= ~ -~\frac{i}{2}~ 
{\hat e} {\hat e}^a_l \left( ~ {\overline \psi} \gamma^l_{}
D^{}_t (\omega) \psi  - {\overline {D^{}_t (\omega) \psi}}
~ \gamma^l_{} \psi \right) \label{TFD3}\\
&&T^{~a}_i ~\equiv~ {\hat e}~ {\hat T}^{~a}_i ~= ~  \frac{i}{2}~ 
{\hat e} {\hat e}^a_i \left( ~ {\overline \psi} \gamma^0_{}
D^{}_t (\omega) \psi  - {\overline {D^{}_t (\omega) \psi}}
~ \gamma^0_{} \psi \right) \label{TFD4}
\end{eqnarray}
We use the solutions (\ref{omegaF1}), (\ref{omegaF2})  and 
(\ref{eqF2}) in these equations:
\begin{eqnarray}
&&T^{~t}_0 ~\equiv~ {\hat e}~ {\hat T}^{~t}_0 ~= ~ {\hat e}
\left[ \frac{i}{2}~ {\hat e}^a_l \left(~ {\overline \psi} \gamma^l_{}
D^{}_a ({\bar \omega}) \psi  - {\overline {D^{}_a ({\bar \omega}) \psi}}
~ \gamma^l_{} \psi \right) ~+~ m {\overline \psi }\psi 
\right.\nonumber \\
&& \left. ~~~~~~~~~~~~~~~~~~~~~~~~~~~~~~~~~~~~~~~+ \frac{1}{2} ~
N^{}_{ll} {\overline \psi} \gamma^{}_5 \gamma^{}_0 \psi 
- \frac{2}{\kappa^2_{}}~ {\tilde M}^{ij}_{} {\tilde M}^{ij}_{} \right] 
\label{TFD5}\\
&&T^{~t}_i ~\equiv~ {\hat e}~ {\hat T}^{~t}_i ~= ~ -~\frac{i}{2}~ 
{\hat e} {\hat e}^a_i \left( ~ {\overline \psi} \gamma^0_{}
D^{}_a ({\bar \omega}) \psi  - {\overline {D^{}_a ({\bar \omega}) 
\psi}}~ \gamma^0_{} \psi + \frac{2i}{\kappa^2{}}~
\epsilon^{lmn}_{} N^{~l}_a M^{mn}_{} \right) \label{TFD6}\\
&&T^{~a}_0 ~\equiv~ {\hat e}~ {\hat T}^{~a}_0 ~= ~ 0 \label{TFD7}\\
&&T^{~a}_i ~\equiv~ {\hat e}~ {\hat T}^{~a}_i ~= ~0 \label{TFD8}
\end{eqnarray}

Now we analyze the last set of sixteen  Euler-Lagrange equations 
contained in (\ref{eqtetradF}). As earlier, we break these into 
four sets of $1$, $3$, $3$ and $9$ equations as:
\begin{eqnarray}
\kappa^2_{}~ T^{~t}_0  ~\equiv~ {\hat e}~ \kappa^2_{}~ 
{\hat T}^{~t}_0 ~&=&~ -~ \frac{1}{2}~{\hat e} 
\left[ {\hat e}^b_k ~{\hat e}^c_l~ R^{~~kl}_{bc} (\omega) 
~-~ 2 \Lambda \right]  \label{eqtetradDF1} \\
\kappa^2_{}~ T^{~t}_i  ~\equiv~ {\hat e}~ \kappa^2_{}~ 
{\hat T}^{~t}_i ~&=&~ {\hat e}~ {\hat e}^b_i~ {\hat e}^c_l 
~ R^{~~0l}_{bc} (\omega) \label{eqtetradDF2}\\
\kappa^2_{}~ T^{~a}_0  ~\equiv~ {\hat e}~ \kappa^2_{}~ 
{\hat T}^{~a}_0 ~&=&~ {\hat e}~ {\hat e}^a_k~ {\hat e}^b_l 
~ R^{~~kl}_{tb} (\omega) \label{eqtetradDF3} \\
\kappa^2_{}~ T^{~a}_i  ~\equiv~ {\hat e}~ \kappa^2_{}~ 
{\hat T}^{~a}_i ~&=&~ {\hat e}~ {\hat e}^a_{[i}~ {\hat e}^b_{j]} 
~ R^{~~0j}_{tb} (\omega) \label{eqtetradDF4}
\end{eqnarray}
where now  various components of $T^{~\mu}_I$ are  given by 
Eqns.(\ref{TFD5}-\ref{TFD8}). The last two equations are exactly the 
same as earlier for the scalar and vector gauge matter field cases
and hence we have:
\begin{eqnarray}
&&{\hat e}^a_i~ R^{~~ij}_{ta} (\omega) ~=~0 \label{solRF1}\\
&& R^{~~0i}_{ta} (\omega)~=~ \partial^{}_t M^{~i}_a ~=~0 
\label{solRF2}
\end{eqnarray}
where we have used $\omega^{~0i}_t ~=~0$ and $\omega^{~ij}_t ~=~0$ 
in the second equation. From Eqns.(\ref{eqtetradDF2}) and 
(\ref{TFD6}), we have:
\begin{eqnarray}
{\hat e}^b_l ~R^{~~0l}_{ab} (\omega) ~&=&~  \kappa^2_{} ~e^i_a ~
{\hat T}^{~t}_i \nonumber \\
~&=&~ - ~ \frac{i}{2}~ \kappa^2_{}~
\left( {\overline \psi} \gamma^0_{} D^{}_a (\omega) \psi 
~-~ {\overline{D^{}_a (\omega) \psi }} ~\gamma^0_{} \psi \right) 
\nonumber \\
&=&~ - ~ \frac{i}{2}~ \kappa^2_{}~
\left( {\overline \psi} \gamma^0_{} D^{}_a ({\bar \omega}) \psi 
~-~ {\overline{D^{}_a ({\bar \omega}) \psi }} ~\gamma^0_{} \psi \right)
~+~ \epsilon^{lmn}_{} N^{~l}_a M^{mn}_{} 
\label{solRF3}
\end{eqnarray}
Note that
\begin{eqnarray*}
{\hat e}^b_l ~R^{~~0l}_{ab} (\omega) ~=~ {\hat e}^b_l~ 
D^{}_{[a} ( \omega) M^{~l}_{b]} &=~ {\hat e}^b_l~ 
D^{}_{[a} ({\bar \omega}) M^{~l}_{b]}  ~
+ \epsilon^{lmn}_{} N^{~l}_a M^{mn}_{}
\end{eqnarray*}
Using this in (\ref{solRF3}), we have
\begin{equation}
{\hat e}^b_l~ D^{}_{[a} ( {\bar \omega}) M^{~l}_{b]} ~=~
 - ~ \frac{i}{2}~ \kappa^2_{}~
\left( {\overline \psi} \gamma^0_{} D^{}_a ({\bar \omega}) \psi 
~-~ {\overline{D^{}_a ({\bar \omega}) \psi }} ~\gamma^0_{} \psi \right)
\label{solRF4}
\end{equation}  
We break $M^{~l}_a$ as $M^{~l}_a = {\bar M}^{~l}_a + 
{\tilde M}^{~l}_a$ where ${\bar M}^{~l}_a \equiv e^m_a 
{\bar M}^{ml}_{}$ and ${\tilde M}^{~l}_a \equiv e^m_a 
{\tilde M}^{ml}_{}$ with ${\bar M}^{ml}_{}$ and ${\tilde M}^{ml}_{}$ 
as the symmetric and antisymmetric parts of the matrix 
$M^{ml}_{}$. For the antisymmetric ${\tilde M}^{ij}_{}$   given 
in terms of the fermions as in (\ref{omegaF2}), it is straight 
forward to check that it satisfies the following equation:
\begin{eqnarray}
{\hat e}^b_l~ D^{}_{[a} ( {\bar \omega}) {\tilde M}^{~l}_{b]} ~=~
 - ~ \frac{\kappa^2_{}}{4{\hat e}}~  g^{}_{ab}~ \epsilon^{bcd}_{}
~ \partial^{}_c \left( {\overline \psi} 
\gamma^{}_5 \gamma^{}_d \psi \right)\label{solRF5}
\end{eqnarray}
where $\gamma^{}_d \equiv e^i_d ~\gamma_i^{}$. Substitute this in
(\ref{solRF4}) to obtain the constraint on the symmetric part 
${\bar M}^{ml}_{}$ as:
\begin{eqnarray}
{\hat e}^b_l~ D^{}_{[a} ( {\bar \omega}) {\bar M}^{~l}_{b]} ~&=&~
   \frac{\kappa^2_{}}{4{\hat e}}~  g^{}_{ab}~ \epsilon^{bcd}_{}
~ \partial^{}_c \left( {\overline \psi} 
\gamma^{}_5 \gamma^{}_d \psi \right)\nonumber \\
&& ~~- ~ \frac{i\kappa^2_{}}{2}~ 
\left( {\overline \psi} \gamma^0_{} D^{}_a ({\bar \omega}) \psi 
~-~ {\overline{D^{}_a ({\bar \omega}) \psi }} ~\gamma^0_{} \psi \right)
\label{solRF6}
\end{eqnarray}

Now we are left to analyze the Euler-Lagrange equation of 
motion (\ref{eqtetradDF1}). This we do in the same manner
as in  earlier cases of scalar and vector gauge field matter
to obtain:
\begin{eqnarray}
{\hat e}^b_k {\hat e}^c_l~ {\bar R} ^{~~kl}_{bc} ({\bar \omega} ) 
&-&  \left( M^{kl}_{} M^{lk}_{} - M^{kk}_{} M^{ll}_{} \right)
+  \left( N^{kl}_{} N^{lk}_{} - N^{kk}_{} N^{ll}_{} \right)
-  2 \Lambda \nonumber \\
&=& -~ 2\kappa^2_{} ~{\hat T}^{~t}_0  \nonumber \\
 &= & -   2\kappa^2_{}  \left[ \frac{i}{2}~ {\hat e}^a_l 
 \left(~ {\overline \psi} \gamma^l_{}
D^{}_a (\omega) \psi  - {\overline {D^{}_a (\omega) \psi}}
~ \gamma^l_{} \psi \right) ~+~ m {\overline \psi }\psi \right]
\nonumber \\ 
&=& -~ 2\kappa^2_{}  \left[ \frac{i}{2}~ {\hat e}^a_l 
\left(~ {\overline \psi} \gamma^l_{}
D^{}_a ({\bar \omega}) \psi  - {\overline {D^{}_a ({\bar \omega}) \psi}}
~ \gamma^l_{} \psi \right) ~+~ m {\overline \psi }\psi 
\right. \nonumber \\
&& ~~~~~~~~~~~~~~~~~~~~~~~~\left. + ~\frac{1}{2} ~ 
N^{}_{ll} {\overline \psi} \gamma^{}_5 \gamma^{}_0 \psi 
- \frac{2}{\kappa^2_{}}~ {\tilde M}^{ij}_{} {\tilde M}^{ij}_{} \right]
    \label{solRF7}
\end{eqnarray}

As in earlier cases, Eqn.({\ref{solRF1}) has no additional information 
beyond that already contained in (\ref{omegaF1}). Thus we are left
with (\ref{solRF2}), (\ref{solRF6}) and (\ref{solRF7}) as the 
set of constraints. A particular solution of the constraints
(\ref{solRF2}) and (\ref{solRF6}) is given by:
\begin{eqnarray}
&& {\bar M}^{~i}_a ~=~ \lambda ~ e^i_a~, ~~~\partial^{}_t \lambda ~=~0~,
\nonumber \\
&&\partial^{}_a \lambda ~=~
\frac{\kappa^2_{}}{8{\hat e}}~  g^{}_{ab}~ \epsilon^{bcd}_{}
~ \partial^{}_c \left( {\overline \psi} 
\gamma^{}_5 \gamma^{}_d \psi \right) 
   -   \frac{i\kappa^2_{}}{4}~ 
\left( {\overline \psi} \gamma^0_{} D^{}_a ({\bar \omega}) \psi 
~-~ {\overline{D^{}_a ({\bar \omega}) \psi }} ~\gamma^0_{} \psi \right)
  \label{solRF8}
\end{eqnarray}
This leaves us with the master constraint (\ref{solRF7}).

Note that, from Eqn.(\ref{solRF7}), we have:
\begin{equation}
\partial^{}_t \xi ~=~ \kappa^2{} ~\partial^{}_t N^{ll}_{} ~
{\overline \psi} \gamma^{}_5 \gamma^{}_0 \psi ~, ~~~~ 
\xi~\equiv~ N^{ll}_{} N^{kk}_{} - N^{lk}_{} N^{kl}_{} \label{txi}
\end{equation}
where we have used the fact that due to equations of motion all but
$N^{ij}_{}$ in Eqn.(\ref{solRF7}) are $t$ independent. This equation 
can also equivalently be  written as:
\begin{eqnarray*}
N^{lm}_{} \partial^{}_t N^{ml}_{}~=~ \left( N^{ll}_{} - 
\frac{\kappa^2_{}}{2}~ {\overline \psi} \gamma^{}_5
\gamma^{}_0 \psi\right) \partial^{}_t N^{kk}
\end{eqnarray*}
Eqn (\ref{txi}) is equivalent to Eqn.(\ref{TF2}) for $I=0$ for 
the degenerate tetrads (\ref{tetradD}). For $I=i$, Eqn.(\ref{TF2})
is identically satisfied for configurations obeying
the equations of motion listed above.

\section{Concluding remarks}

We have extended the discussion of degenerate metrics in first order 
gravity by including matter fields. This has been done using
first order action functionals for the matter fields. Like the
Hilbert-Palatini action, these matter actions are defined for both
invertible and non-invertible tetrads. The Euler-Lagrange equations 
of motion obtained from so constructed actions are also defined for
invertible as well as non-invertible metrics. This provides an 
appropriate framework to study degenerate metrics. The phase 
containing non-degenerate tetrads  provides a description which 
is equivalent to second order theory of gravity coupled to matter. 
For degenerate metrics, the theory exhibits a new phase with very 
different structure which has been displayed through a  detail 
analysis for non-invertible tetrads with one zero eigen value.
The matter fields considered are scalar, $U(1)$ vector gauge fields 
and fermions. Generalization to other matter fields like non-Abelian 
vector gauge fields is straight forward and can be done in a similar 
manner. Also this analysis can be extended to study degenerate 
tetrads with more that one zero eigen values in a similar
spirit.

\acknowledgments
Discussions with Sandipan Sengupta  and financial support from 
the Department of Atomic Energy, Government of India are 
gratefully acknowledged.

\end{document}